\documentclass[conference,draftcls,onecolumn,12pt]{IEEEtranTCOM}
\usepackage{amsmath, amsthm, amsfonts, amssymb, amsbsy}

\usepackage{graphicx}
\usepackage{float}
\usepackage{algorithm}
\usepackage{algorithmic}
\usepackage{multirow}
\usepackage{bm}
\usepackage{stfloats}
\usepackage{subfigure}
\usepackage[usenames]{color}
\usepackage{colortbl,booktabs}
\definecolor{cA}{rgb}{.36,.61,.84}
\definecolor{cB}{rgb}{.44,.68,.28}
\definecolor{cC}{rgb}{.5,.5,.5}
\definecolor{cD}{rgb}{.77,.35,.07}

\newtheorem{example}{Example}
\newcommand{\eax}{\twotriangle  \end{example}}

\newcommand{\ba}{{\bf a}}
\newcommand{\bb}{{\bf b}}
\newcommand{\bc}{{\bf c}}
\newcommand{\bd}{{\bf d}}
\newcommand{\be}{{\bf e}}
\newcommand{\bp}{{\bf p}}
\newcommand{\bq}{{\bf q}}
\newcommand{\bu}{{\bf u}}
\newcommand{\bv}{{\bf v}}

\newcommand{\bM}{{\bf M}}

\title{
Generalized Piggybacking Codes for Distributed Storage Systems
}

\IEEEoverridecommandlockouts

\begin{document}

\author{
Shuai Yuan$^1$, Qin Huang$^{2,1,*}$, \emph{Senior Member, IEEE}, Zulin Wang$^2$, \emph{Member, IEEE}\\
$^1$Qian Xuesen Laboratory of Space Technology\\
China Academy of Space Technology, Beijing, China, 100094\\
$^2$School of Electronic and Information Engineering\\
Beihang University, Beijing, China, 100191\\
Email: yuanshuai@qxslab.cn; qhuang.smash@gmail.com; wzulin\_201@163.com\\
\thanks{Part of this paper has been accepted by IEEE Global Communications Conference (IEEE Globecom 2016).
	
	Corresponding author: Q. Huang.}
}

\markboth{IEEE Transactions on Communications}%
{Submitted paper}

\maketitle

\begin{abstract}
 This paper generalizes the piggybacking constructions for distributed storage systems by considering various protected instances and piggybacked instances. Analysis demonstrates that the proportion of protected instances determines the average repair bandwidth for a systematic node. By optimizing the proportion of protected instances, the repair ratio of generalized piggybacking codes approaches zero instead of 50\% as the number of parity check nodes tends to infinity. Furthermore,  the computational complexity for repairing a single systematic node cost by generalized piggybacking codes is less than that of the existing piggybacking designs. 
\end{abstract}

\begin{IEEEkeywords}
piggybacking, distributed storage systems, MDS, node repair
\end{IEEEkeywords}

\section{Introduction}

Nowadays, distributed storage systems (DSSs) are being increasingly employed by network applications. Data in DSSs is deployed over multiple storage devices. However, these discrete devices are prone to failure because of malfunctions or maintenance. In order to ensure the reliability of the stored data even in the occurrence of node unavailability, DSSs are supposed to introduce redundancy to resist storage node failures. Replication is the simplest redundant fashion, and has been adopted to improve the reliability by many DSSs, such as the Google File System \cite{gfs2003} and the Hadoop Distributed File System (HDFS) \cite{hdfs2008}. With the rapid growth of amount of storage data, erasure coding has become a better choice for DSSs. Compared with replication, it is able to provide orders of magnitude reliability increasing for same storage resource consumption \cite{erasurecoding2002}. As a result, several large-scale systems, such as OceanStore \cite{oceanstore2003}, Total Recall \cite{totalrecall2004}, Windows Azure Storage \cite{was2011}, and Google Colossus(GFS2) \cite{gfs22012}, have employed erasure coding techniques to improve their storage efficiency.

\emph{Maximum distance separable} (MDS) codes as one kind of erasure codes have been introduced into many DSSs for their optimal storage efficiency. MDS property can be used to recover missing data in a DSS. Consider an $n$-node DSS deployed with an $(n,k)$ MDS code. If one node of this storage system is failed, data stored in $k$ nodes is required to reconstruct the missing data in this failure node. $k$ times amount of stored data is needed to recover the missing data. Thus, the usage of network and disk is significantly high, i.e., the repair efficiency is very low. To address this repair issue, many codes have been constructed to reduce the transmission data for repairing failure node.

As the statement in \cite{dimakis2011survey}, there are three types of node repair: exact repair, functional repair and exact repair of the systematic part. However, exact repair is the most considered from in practical DSSs. In \cite{regeneratingcode2010}, Dimakis et al. defined the amount of transmission data during repairing  one single failed node as \emph{repair bandwidth}. The authors derived an optimal tradeoff between storage and repair bandwidth (theoretic cut-set bound), and proposed \emph{regenerating codes} which lie on the tradeoff curve. In \cite{searching2009,rashmi2009explicit,interalign2010ISIT,pm2011rashmi,li2015regenerating}, the existence and the construction of regenerating codes have been studied. However, the optimal tradeoff provided by regenerating codes was only derived for functional repair. Almost all the interior points on the storage-bandwidth tradeoff are not achievable under exact repair \cite{nonachieveability2012shah}.


MDS array codes are another important class of erasure codes used in DDSs. They have the advantage of simple encoding and decoding procedures, so that they can be easily implemented in hardware devices. Many designs of MDS array codes, such as EVENODD \cite{blaum1995evenodd}, B-code \cite{xu1999B-code}, X-code  \cite{xu1999x-code}, RDP \cite{corbett2004rdp}, STAR \cite{huang2008star} and Zigzag codes \cite{tamo2013zigzag}, have been presented for storage and communication applications. However, the repair bandwidth of MDS array codes can not achieve the theoretic cut-set bound.

In 2011, Rashmi et al. proposed a new kind of distributed storage codes called \emph{piggybacking codes} to reduce the data amount read and downloaded for node repair \cite{piggyback2011ISIT}. The key idea of piggybacking codes is taking several instances of an existing base code, and attaching linear combinations of symbols in some protected instances to other non-protected instances. Hence, the missing symbols in protected instances are able to be recovered by solving these linear equations instead of MDS decoding. Piggybacking is a simple and useful construction to improve the repair efficiency of missing nodes. Several designs of piggybacking codes were presented in \cite{piggyback2011ISIT} and \cite{piggyback2013arxiv}. These designs are able to save $25\%$ to $50\%$ repair bandwidth for one failed node on average. Facebook Warehouse Cluster and the new Hadoop Distributed File System (HDFS) have employed piggybacking codes to improve their repair efficiency \cite{facebook2013}.

Although piggybacking codes are practical and easy to implementation, the reduction of repair bandwidth of the proposed piggybacking designs still has a gap to the theoretic cut-set bound of regenerating codes. In \cite{piggyback2013arxiv}, Rashmi, Shah, and Ramchandran gave three specific piggybacking constructions.  
The second one we represent with RSR-II is the most efficient construction in terms of repair bandwidth. The description in \cite{piggyback2013arxiv} shows that RSR-II codes are able to save up to $50\%$ of repair bandwidth. This paper investigates the mechanism in reduction of repair bandwidth by using piggybacking codes. From the recovery methods of the systematic symbols, we distinguish instances of piggybacking codes with protected stripes and non-protected stripes. An analysis of a lower bound on the repair bandwidth of RSR-II codes implies that the proportion of protected instances determines the repair efficiency of piggybacking constructions.

This paper firstly presents a generalized piggybacking design with various protected and non-protected stripes in order to obtain various proportion of protected stripes. Second, a lower bound and an upper bound on the repair bandwidth of generalized piggybacking codes are introduced. The analysis of the two bounds indicates that by optimizing the proportion of protected stripes, the \emph{repair ratio} ( defined as average repair bandwidth as a fraction of the amount of original messages) of a generalized piggybacking code approaches zero instead of $50\%$ as the number of parity check nodes tends to infinity. It is closer to that of minimum storage regenerating (MSR) codes which has the theoretical lower bound. At last, the computational complexity for the repair of a single failed systematic node is analyzed. The results show that the generalized piggybacking codes are able to provide more efficient repair with little complexity overhead.

The remainder of this paper is organized as follows. Section II briefly introduces the piggybacking framework and RSR-II codes. Section III performs an analysis of the repair efficiency of RSR-II codes. Our generalized piggybacking codes are presented in Section IV. Finally, the conclusion is given in Section V.

\section{Background}
\subsection{Maximum distance separable codes}\label{sec.mds}

Consider an $(n,k,d)$ linear block code $\cal{C}$, where $n$ is its code length, $k$ is its dimension, and $d$ represents the minimum Hamming distance. Code $\cal{C}$ is called an MDS code, if its minimum Hamming distance $d$ meets the Singleton bound, i.e.,
\begin{equation}
d=n-k+1.
\end{equation}

MDS codes are an important class of linear block codes. For given parameters $n$ and $k$, the minimum distance $d$ reaches the maximum possible value. Thus, MDS codes are able to correct as many as $(n-k)$ erasures for given $n$ and $k$.

MDS codes have been extensively applied in many DSSs. In an $n$-node storage system, initially the original message is divided into $k$ information packets. Subsequently, the $k$ packets are encoded into $n$ packets and stored in the $n$ nodes respectively. With the MDS property, messages from any $k$ out of $n$ nodes could reconstruct the original message. Thus, the system is able to tolerate the failures of any $(n-k)$ storage nodes.

\subsection{Piggybacking framework}\label{sec.piggyfrm}

In this subsection, we introduce the piggybacking framework which is the basis of constructing piggybacking codes. Piggybacking framework guarantees that DSSs are able to employ piggybacking codes without extra cost of storage. Moreover, the decoding properties of the error-correction codes adopted by original DSSs, such as the minimum distance or the MDS property, are not ruined by piggybacking reconstruction.

In general, the piggybacking framework operates on multiple instances of an existing base code and adds several designed functions of the data in some instances onto other instances. The base code of piggybacking framework can be arbitrary. In fact, it is a very attractive feature in practice. Under the piggybacking framework, the DSSs enjoy a repair bandwidth reduction with only small modification based on their existing error-correction codes.

Consider a linear block code ${\cal C}_1$ represented by $n$ encoding functions $\{f_i\}^n_{i=1}$. Suppose $\bu$ is the original message of ${\cal C}_1$. The $n$ encoded symbols are $\{f_i(\bu)\}_{i=1}^n$. For an $n$-node system, using ${\cal C}_1$ as the base code, the piggybacking framework, which has $\alpha$ instances of ${\cal C}_1$, is illustrated in Fig.\ref{fig.pbframe}.

\begin{figure*}[t]
	\begin{center}
		\small
		\begin{tabular}{c c c c c c }
			& stripe $1$ & stripe $2$ & stripe $3$ & $\cdots$ & stripe $\alpha$\\
			\cline{2-6}
			\multicolumn{1}{l|}{node $1$} & \multicolumn{1}{c|}{$f_1({\bu}_1)$} & \multicolumn{1}{c|}{$f_1({\bu}_2)+g_{2,1}({\bu}_1)$} & \multicolumn{1}{c|}{$f_1({\bu}_3)+g_{3,1}({\bu}_1,{\bu}_2)$} & \multicolumn{1}{c|}{$\cdots$} & \multicolumn{1}{c|}{$f_1({\bu}_{\alpha})+g_{\alpha,1}({\bu}_1,\cdots,{\bu}_{\alpha-1})$}\\
			\cline{2-6}
			\multicolumn{1}{l|}{node $2$} & \multicolumn{1}{c|}{$f_2({\bu}_1)$} & \multicolumn{1}{c|}{$f_2({\bu}_2)+g_{2,2}({\bu}_1)$} & \multicolumn{1}{c|}{$f_2({\bu}_3)+g_{3,2}({\bu}_1,{\bu}_2)$} & \multicolumn{1}{c|}{$\cdots$} & \multicolumn{1}{c|}{$f_2({\bu}_{\alpha})+g_{\alpha,2}({\bu}_1,\cdots,{\bu}_{\alpha-1})$}\\
			\cline{2-6}
			\multicolumn{1}{l|}{$\quad\vdots$} & \multicolumn{1}{c|}{$\vdots$}    & \multicolumn{1}{c|}{$\vdots$}   & \multicolumn{1}{c|}{$\vdots$}      & \multicolumn{1}{c|}{$\ddots$} & \multicolumn{1}{c|}{$\vdots$}\\
			\cline{2-6}
			\multicolumn{1}{l|}{node $n$} & \multicolumn{1}{c|}{$f_n({\bu}_1)$} & \multicolumn{1}{c|}{$f_n({\bu}_2)+g_{2,n}({\bu}_1)$} & \multicolumn{1}{c|}{$f_n({\bu}_3)+g_{3,n}({\bu}_1,{\bu}_2)$} & \multicolumn{1}{c|}{$\cdots$} & \multicolumn{1}{c|}{$f_n({\bu}_{\alpha})+g_{\alpha,n}({\bu}_1,\cdots,{\bu}_{\alpha-1})$}\\
			\cline{2-6}
		\end{tabular}
	\end{center}
	\caption{Piggybacking framework}\label{fig.pbframe}
\end{figure*}

As shown in Fig.\ref{fig.pbframe}, the $n$ rows correspond to the $n$ storage nodes, the $\alpha$ columns are called $\alpha$ \emph{stripes}, $\{{\bu}_i\}_{i=1}^\alpha$ are $\alpha$ independent original messages and $\{g_{i,j}\}_{i=2,j=1}^{\alpha,n}$ are \emph{piggyback functions}. 

It is a very important consideration that the piggyback functions added on the $i$-th stripe $(i\in\{2,3,\cdots,\alpha\})$ can only be linear combinations of original messages of stripes $\{1,2,\cdots,(i-1)\}$. This principle guarantees that all the stripes of this piggybacking framework are decodable through a recursion process: In stripe 1, no piggyback functions are added, so the original message ${\bu}_1$ can be directly recovered by using the decoding procedure of ${\cal C}_1$. For stripe 2, with the decoded ${\bu}_1$, it is easy to compute the added piggyback functions $\{g_{2,j}({\bu}_1)\}_{j=1}^n$ and subtract them from the stored symbols. Then, ${\bu}_2$ is decodable. In a similar way, after the decoding procedures of stripes $\{1,2,\cdots,(i-1)\}$ are finished, ${\bu}_1,{\bu}_2,\cdots,{\bu}_{i-1}$ are available to the piggyback functions $\{g_{i,j}({\bu}_1,\cdots,{\bu}_{i-1})\}_{j=1}^n$. The base code of this stripe is obtained after subtracting these piggybacking functions, so that ${\bu}_{i}$ can be recovered.

As the statement above, the $\alpha$ symbols stored in one node are independent. Sometimes, an invertible linear transformation is performed to simplify the computation. Such a transformation still retains the decoding properties of the piggybacking framework.

\subsection{RSR-II codes}\label{sec.rsrcode}

Under the piggybacking framework described in Section.\ref{sec.piggyfrm}, Rashmi et al. have presented three designs of piggybacking codes for different considerations. The second design RSR-II is constructed for the purpose of pursuing high efficiency of repair. As the statement in \cite{piggyback2013arxiv}, RSR-II codes can save up to $50\%$ repair bandwidth of a systematic node.

For the sake of simple description, an $(n,k)$ MDS code in systematic form is chosen as the base code. Denote $r=n-k$ as the number of parity check nodes. RSR-II codes consist of $(2r-3)$ instances of the base code. Represent the $(2r-3)$ associated original messages as $\ba_1,\ba_2,\cdots,\ba_{2r-3}$, where $\ba_i$ $(i\in\{1,2,\cdots,2r-3\})$ is a vector of length $k$, and $\ba_i= [a_{i,1},a_{i,2},\cdots,a_{i,k}]$. Then, the $(2r-3)$ stripes are shown in the following form:
\begin{center}
	\small
	\begin{tabular}{l|c|c|c|c|}
		\cline{2-5}
		node $1$ & $a_{1,1}$ & $a_{2,1}$ & $\cdots$ & $a_{2r-3,1}$\\
		\cline{2-5}
		$\quad\vdots$ & $\vdots$      & $\vdots$      & $\ddots$ & $\vdots$\\
		\cline{2-5}
		node $k$ & $a_{1,k}$ & $a_{2,k}$ & $\cdots$ & $a_{2r-3,k}$\\
		\cline{2-5}
		node $k$+$1$ & $\bp^T_1\ba_1$ & $\bp^T_1\ba_2$ & $\cdots$ & $\bp^T_1\ba_{2r-3}$\\
		\cline{2-5}
		$\quad\vdots$ & $\vdots$      & $\vdots$      & $\ddots$ & $\vdots$\\
		\cline{2-5}
		node $k$+$r$ & $\bp^T_r\ba_1$ & $\bp^T_r\ba_2$ & $\cdots$ & $\bp^T_r\ba_{2r-3}$\\
		\cline{2-5}
	\end{tabular}
\end{center}
where $\bp_1,\bp_2,\cdots,\bp_r$ are $r$ encoding vectors corresponding to the $r$ parity check symbols of the base code.

The piggyback functions of RSR-II codes are $(r-1)^2$ linear combinations of the systematic symbols of the first $(r-1)$ stripes, and they are added on the last $(r-1)$ parity check symbols of the last $(r-1)$ stripes. The construction of these piggyback functions is taken in three steps.

First, the $k$ systematic nodes are split into $(r-1)$ node sets $\{S_i\}_{i=1}^{r-1}$ as evenly as possible. Without loss of generality, we suppose $k$ is not a multiple of $(r-1)$, and define three variables as follows,
\begin{equation}\label{equ.size1}
	t_l=\Bigl\lfloor \frac{k}{r-1}\Bigr\rfloor,\ t_h=\Bigl\lceil \frac{k}{r-1}\Bigr\rceil,\ t=k-(r-1)t_l.
\end{equation}
Hence, the first $t$ node sets $\{S_i\}_{i=1}^t$ are of size $t_h$, and the remaining $\{S_i\}_{i=t+1}^{r-1}$ are of size $t_l$.

Second, define two sets of vectors of length $k$ $\{\bv_i\}_{i=2}^r$ and $\{\hat{\bv}_i\}_{i=2}^r$ with
\begin{eqnarray}
	\bv_i&=&\ba_{r-1}+i\ba_{r-2}+i^2\ba_{r-3}+\cdots+i^{r-2}\ba_1,  \\
	\hat{\bv}_i&=&\bv_i-\ba_{r-1}=i\ba_{r-2}+i^2\ba_{r-3}+\cdots+i^{r-2}\ba_1.
\end{eqnarray}
Then, introduce $(r-1)^2$ selection vectors $\{\bq_{i,j}\}_{i=2,j=1}^{r,r-1}$ to separate the $k$ tuples in each vector of $\{\bv_i\}_{i=2}^r,\{\hat{\bv}_i\}_{i=2}^r$ into $(r-1)$ segments. And the selection vectors are defined as follows
\begin{equation}
	\bq_{i,j} = \bM_j\bp_i,
\end{equation}
where $\{\bM_j\}_{j=1}^{r-1}$'s are diagonal matrices of size $(k\times k)$. On the diagonal of $\bM_j$, only the positions corresponding to the systematic nodes in $S_j$ are ``1''. Therefore,
\begin{equation}
	\sum\limits_{j=1}^{r-1}\bq_{i,j}=\bp_i,\ \forall i\in\{2,\cdots,r\}.
\end{equation}

Finally, add the piggyback functions of $\{\bv_i\}_{i=2}^r,\{\hat{\bv}_i\}_{i=2}^r$ and $\{\bq_{i.j}\}_{i=2,j=1}^{r,r-1}$ into the parity check symbols in the last $(r-1)$ nodes. Hence, node $(k$+$i)$, $i\in\{2,3,\cdots,r\}$, has the following form as shown in Fig.\ref{fig.nodesymbol_1}. An invertible linear transformation is introduced to reduce the complexity for node repair. Finally, symbols in node $(k$+$i)$ are illustrated in Fig.\ref{fig.nodesymbol_2}.

\begin{figure*}
	\begin{center}
		\subfigure[node $(k$+$i)$ with piggyback functions]{\label{fig.nodesymbol_1}
		\small
		\begin{tabular}{|c|c|c|p{2.4cm}|l|c|l|l|c|l|}
			\hline
			\multirow{2}{*}{$\bp_i^T \ba_1$}&\multirow{2}{*}{$\cdots$}&\multirow{2}{*}{$\bp_i^T \ba_{r-2}$}&$\bp_i^T \ba_{r-1}+$&$\bp_i^T \ba_r+$&\multirow{2}{*}{$\cdots$}&$\bp_i^T \ba_{r+i-3}+$&$\bp_i^T \ba_{r+i-2}+$&\multirow{2}{*}{$\cdots$}&$\bp_i^T \ba_{2r-3}+$\\
			&&&$\sum_{j=1,j\ne i-1}^{r-1}\bq_{i,j}^T \hat{\bv}_i$&$\bq_{i,1}^T \bv_i$&&$\bq_{i,i-2}^T \bv_i$&$\bq_{i,i}^T \bv_i$&&$\bq_{i,r-1}^T \bv_i$\\
			\hline
		\end{tabular}}
		\subfigure[node $(k$+$i)$ with an invertible linear transform]{\label{fig.nodesymbol_2}			
		\small
		\begin{tabular}{|c|c|c|p{2.4cm}|l|c|l|l|c|l|}
			\hline
			\multirow{2}{*}{$\bp_i^T \ba_1$}&\multirow{2}{*}{$\cdots$}&\multirow{2}{*}{$\bp_i^T \ba_{r-2}$}&$\bq_{i,i-1}^T \ba_{r-1}-$&$\bp_i^T \ba_r+$&\multirow{2}{*}{$\cdots$}&$\bp_i^T \ba_{r+i-3}+$&$\bp_i^T \ba_{r+i-2}+$&\multirow{2}{*}{$\cdots$}&$\bp_i^T \ba_{2r-3}+$\\
			&&&$\sum_{j=r}^{2r-3}\bp_i^T \ba_j$&$\bq_{i,1}^T \bv_i$&&$\bq_{i,i-2}^T \bv_i$&$\bq_{i,i}^T \bv_i$&&$\bq_{i,r-1}^T \bv_i$\\
			\hline
		\end{tabular}}
	\end{center}
	\caption{Stored symbols in piggybacked node $(k$+$i)$}
\end{figure*}

\subsection{Repair bandwidth of RSR-II codes}\label{sec.rb_rsr}

We use \emph{repair ratio} $\gamma$ to represent the measure of repair efficiency of a distributed storage code. Repair ratio is defined as the average amount of transfer data needed for repairing one failure node as a fraction of original messages. In this subsection, we recall the repair procedure of one systematic node by RSR-II codes. Then, the repair ratio of RSR-II $\gamma^{sys}_1$ is computed.

Consider an $n$-node DSS deployed with an $(n,k)$ RSR-II code. For the sake of simple description, we represent the first $(r-1)$ stripes as \emph{protected stripes}, whose systematic symbols are involved in the piggyback functions and defined as \emph{protected symbols}. Meanwhile, the last $(r-2)$ stripes are represented as \emph{non-protected stripes}, whose systematic symbols are named with \emph{non-protected symbols}. If the $l$-th systematic node fails, repair procedure of this node is to recover the missing protected symbols $\{a_{i,l}\}_{i=1}^{r-1}$ and the missing non-protected symbols $\{a_{i,l}\}_{i=r}^{2r-3}$. Assume node $l$ belongs to $S_j$ which is one of the $(r-1)$ node sets described in Section.\ref{sec.rsrcode}. The repair procedure is described in Algorithm \ref{alg.rsr-ii}.

\begin{algorithm}[htb]
	\caption{The repair algorithm of RSR-II codes} \label{alg.rsr-ii}
		\begin{itemize}
		\item[1] Recovering the missing non-protected symbols $\{a_{i,l}\}_{i=r}^{2r-3}$;\\
			The base code of this RSR-II code is in systematic MDS form. According to MDS property, $a_{i,l}$ can be directly recovered with $a_{i,1},\cdots,a_{i,l-1},a_{i,l+1},\cdots,a_{i,k}, {\bp_1^T\ba_i}$.
		\item[2] Getting the piggyback functions involved with the missing protected systems $\{a_{i,l}\}_{i=1}^{r-1}$;\\
			As statement in \ref{sec.rsrcode}, there are $(r-1)$ piggyback functions containing $\bq_{i,j}$'s $(i=[2,\cdots,r])$. These piggyback functions are linear combinations of the protected symbols in $S_j$. Download the $(r-1)$ parity check symbols containing the $(r-1)$ piggyback functions, and subtract the items about $\{\ba_j\}_{j=r}^{2r-3}$. Then, the $(r-1)$ piggyback functions involved with $\{a_{i,l}\}_{i=1}^{r-1}$ are left.
		\item[3] Recovering the missing protected symbols $\{a_{i,l}\}_{i=1}^{r-1}$;\\
			Including $\{a_{i,l}\}_{i=1}^{r-1}$, the other surviving protected symbols in $S_j\backslash l$ are also involved with the $(r-1)$ piggyback functions obtained in step 2. Download these surviving symbols, and subtract them out from the $(r-1)$ piggyback functions. Then, $\{a_{i,l}\}_{i=1}^{r-1}$ can be reconstructed by solving the left $(r-1)$ linear combinations.
		\end{itemize}
\end{algorithm}

From Algorithm \ref{alg.rsr-ii}, $(r-2)k$ symbols are needed to be downloaded in step 1, and $(r-1)$ symbols are needed in step 2. In step 3, if the size of $S_j$ is $t_h$, the number of downloaded symbols is $(r-1)(t_h-1)$. Otherwise, if the size is $t_l$, $(r-1)(t_l-1)$ symbols are downloaded. We denote the average repair bandwidth of one systematic node as $B^{sys}_1$. The number of systematic nodes in the node sets of size $t_h$ is $t\cdot t_h$, and the number of those systematic nodes in the node set of size $t_l$ is $(r-1-t)t_l$. Thus
\begin{eqnarray}
B^{sys}_1&=&\frac{1}{k}[tt_h((r-2)k+(r-1)t_h)\nonumber\\
&&+(r-1-t)t_l((r-2)k+(r-1)t_l)].
\end{eqnarray}
Thus, the repair ratio $\gamma^{sys}_1$ is
\begin{eqnarray}\label{equ.gamma1}
\gamma^{sys}_1&=&\frac{B^{sys}_1}{k(2r-3)}\nonumber\\
&=&\frac{1}{k^2(2r-3)}[tt_h((r-2)k+(r-1)t_h)+ \nonumber\\
&&(r-1-t)t_l((r-2)k+(r-1)t_l)] \nonumber\\
&=&\frac{1}{k^2(2r-3)}[k^2(r-2)+ \nonumber\\
&&(tt_h^2+(r-1-t)t_l^2)(r-1)].
\end{eqnarray}

\section{Efficiency Analysis for RSR-II Codes}\label{sec.analysis_rsrii}

In this section, a further analysis on the repair efficiency of RSR-II is performed. 

Here, we introduce a notation \emph{stripe-repair ratio} $\eta$ to measure the repair efficiency of one stripe
\begin{equation}
\eta\triangleq\frac{\text{repair bandwidth for a systematic symbol}}{\text{the amount of original message of this stripe}}.\nonumber
\end{equation}
Consider a piggybacking code with $\beta$ stripes. Assume the stripe-repair ratios of these stripes are $\{\eta_i\}_{i=1}^\beta$. Denote the proportions of these stripes as $\{p_i\}_{i=1}^\beta$. Thus, the repair ratio for systematic nodes of this piggybacking code $\gamma^{sys}$ has the following form,
\begin{equation}\label{equ.sysratio}
	\gamma^{sys} = \sum\limits_{i=1}^{\beta}{p_i\eta_i}.
\end{equation}

Recall the RSR-II codes described in Section \ref{sec.rb_rsr}. The repair procedure deals with the missing protected and non-protected symbols in two different measures: MDS decoding is adopted for the recovery of non-protected symbols, and the amount of downloading for repairing one missing non-protected symbol is $k$ symbols. As regard to the missing protected symbols, solving linear combinations is employed, and the average bandwidth is $t_h$ or $t_l$, which depends on the size of node set containing the failure node. Denote $\eta_p$ and $\eta_{np}$ as the stripe-repair ratios of protected and non-protected stripes, respectively. The amount of original message of one stripe equals to the $k$ symbols stored in the systematic nodes. Hence,
\begin{eqnarray}
	\eta_p&\approx&\frac{t_h\text{ or }t_l}{k}\approx\frac{1}{r-1}\label{equ.etap}\\
	\eta_{np}&=&1.
\end{eqnarray}
Although only an approximate value of $\eta_p$ is given by Equation (\ref{equ.etap}), it is obvious that $\eta_p < \eta_{np}$, i.e., repair procedure for protected stripes requires less downloaded symbols compared with non-protected stripes. This is the mechanism in reduction of repair bandwidth by using piggybacking codes.

In the remainder of this section, we explore the critical factors influencing the repair efficiency through an analysis of $\gamma^{sys}_1$. Represent the proportion of protected stripes with $p_p$. Thus, the proportion of non-protected stripes is $(1-p_p)$. Rewrite $\gamma^{sys}_1$ as the form of Equation (\ref{equ.sysratio}). Then,
\begin{eqnarray}\label{equ.aly_gamma1}
	\gamma^{sys}_1&=& \frac{r-2}{2r-3}\cdot\frac{k^2}{k^2}+\frac{r-1}{2r-3}\cdot\frac{tt_h^2+(r-1-t)t_l^2}{k^2}\nonumber\\
	&=&(1-p_p)\cdot\eta_{np}+p_p\cdot\eta_p,
\end{eqnarray}
where $p_p=\frac{r-1}{2r-3}$, $\eta_{np}=1$ and $\eta_p = \frac{tt_h^2+(r-1-t)t_l^2}{k^2}$. The inequality of quadratic and arithmetic means tells that for $x$ nonnegative integers $n_1,n_2,\cdots,n_x$, they satisfy the following inequality.
\begin{eqnarray}
\sum\limits_{i=1}^{x}n_i^2\ge\frac{\left(\sum\limits_{i=1}^{x}n_i\right)^2}{x}.
\end{eqnarray}
Thus,
\begin{eqnarray}
	\eta_p&=& \frac{tt_h^2+(r-1-t)t_l^2}{k^2}\nonumber\\
	&\ge&\frac{(tt_h+(r-1-t)t_l)^2}{k^2(r-1)}\nonumber\\
	&=&\frac{1}{r-1},
\end{eqnarray}
with equality if and only if $t_l=t_h$, i.e., $k$ is a multiple of $(r-1)$. In this case, $\gamma^{sys}_1$ is able to reach a lower bound $\min{(\gamma^{sys}_1)}$, and
\begin{eqnarray}\label{equ.lowerbandgamma1}
\min{(\gamma^{sys}_1)}&=&\frac{r-2}{2r-3} + \frac{r-1}{2r-3}\cdot\frac{1}{r-1} \nonumber\\
&=&\frac{r-1}{2r-3}.
\end{eqnarray}

According to Equation (\ref{equ.lowerbandgamma1}), $\gamma^{sys}_1$ approaches $0.5$ as the number of parity check nodes tends to infinite, i.e., RSR-II codes are able to save at most $50\%$ repair bandwidth. For a DSS whose parameters $(n,k,r)$ are given, in order to further improve the repair efficiency, the structure of piggybacking design is supposed to be modified. As the analysis above, the protected stripe-repair ratio $\eta_p$ is smaller than $\eta_{np}$. It implies that the repair efficiency of piggybacking codes may be improved by increasing $p_p$ according to Equation (\ref{equ.aly_gamma1}). Actually, larger $p_p$ means more protected symbols involved in one piggyback function that leads to the reduction of $\eta_p$. Therefore, it is possible to improve the repair efficiency of piggybacking codes by optimizing the proportion of protected stripes $p_p$.

\section{Generalized Piggybacking Codes}

In this section, we present a generalized construction which contains various protected and non-protected stripes. An analysis is performed to clarify the relationship between repair ratio $\gamma$ and the proportion of protected stripes $p_p$. The results show that our proposed generalized piggybacking codes are able to provide more efficient node repair by optimizing $p_p$. The repair ratio $\gamma^{sys}_2$ of the generalized piggybacking codes approaches zero when the number of the parity check nodes tends to infinity. 

\subsection{Code design}\label{sec.desgn_generalized}

Similarly, choose an $(n,k)$ systematic MDS code ${\cal C}_2$ as the base code of a generalized piggybacking code. $r=n-k$ is the parity check number. Two parameters $s$ and $p$ are introduced to represent the numbers of protected and piggybacked stripes, respectively. Figure \ref{fig.mul_stps} depicts the $(s+p)$ instances of ${\cal C}_2$.

\begin{figure}
	\begin{center}
		\small
		\begin{tabular}{l|p{0.5cm} p{0.25cm} p{0.6cm} |p{0.9cm} p{0.25cm} p{1.0cm}|}
			\cline{2-7}
			node $1$ & \cellcolor{cA}{$a_{1,1}$} & \cellcolor{cA}{$\cdots$} & \cellcolor{cA}{$a_{s,1}$} & \cellcolor{cB}{$a_{s+1,1}$} & \cellcolor{cB}{$\cdots$} & \cellcolor{cB}{$a_{s+p,1}$}\\
			$\qquad\vdots$ & \cellcolor{cA}{$\quad\vdots$}  & \cellcolor{cA}{$\ddots$} & \cellcolor{cA}{$\quad\vdots$} & \cellcolor{cB}{$\quad\vdots$} & \cellcolor{cB}{$\ddots$} & \cellcolor{cB}{$\quad\vdots$}\\
			node $k$ & \cellcolor{cA}{$a_{1,k}$} & \cellcolor{cA}{$\cdots$} & \cellcolor{cA}{$a_{s,k}$} & \cellcolor{cB}{$a_{s+1,k}$} & \cellcolor{cB}{$\cdots$} & \cellcolor{cB}{$a_{s+p,k}$}\\
			\cline{2-4}
			node $k$+$1$ & \cellcolor{cC}{$\bp^T_1\ba_1$} & \cellcolor{cC}{$\cdots$} & \cellcolor{cC}{$\bp^T_1\ba_s$} & \cellcolor{cB}{$\bp^T_1\ba_{s+1}$} & \cellcolor{cB}{$\cdots$} & \cellcolor{cB}{$\bp^T_1\ba_{s+p}$}\\
			\cline{5-7}
			node $k$+$2$ & \cellcolor{cC}{$\bp^T_2\ba_1$} & \cellcolor{cC}{$\cdots$} & \cellcolor{cC}{$\bp^T_2\ba_s$} & \cellcolor{cD}{$\bp^T_2\ba_{s+1}$} & \cellcolor{cD}{$\cdots$} & \cellcolor{cD}{$\bp^T_2\ba_{s+p}$}\\
			$\qquad\vdots$ & \cellcolor{cC}{$\quad\vdots$} & \cellcolor{cC}{$\ddots$} & \cellcolor{cC}{$\quad\vdots$} & \cellcolor{cD}{$\quad\vdots$} & \cellcolor{cD}{$\ddots$} & \cellcolor{cD}{$\quad\vdots$}\\
			node $k$+$r$ & \cellcolor{cC}{$\bp^T_r\ba_1$} & \cellcolor{cC}{$\cdots$} & \cellcolor{cC}{$\bp^T_r\ba_s$} & \cellcolor{cD}{$\bp^T_r\ba_{s+1}$} & \cellcolor{cD}{$\cdots$} & \cellcolor{cD}{$\bp^T_r\ba_{s+p}$}\\
			\cline{2-7}
		\end{tabular}
	\end{center}
	\caption{$(s+p)$ instances of the base code.}\label{fig.mul_stps}
\end{figure}
According to the construction principle of piggybacking framework, piggyback functions added on the $i$-th stripe should only involve the original messages of the stripes $[1,\cdots,i-1]$. For the sake of simple analysis, we add the piggyback functions only on the parity check symbols in non-protected stripes. Redefine the non-protected stripes as piggybacked stripes. As illustrated in Fig.\ref{fig.mul_stps}, all symbols stored in the $(s+p)$ stripes are divided into 4 regions.
\begin{itemize}
	\item Region A contains all the systematic symbols of the protected stripes.
	\item Region B contains all the systematic symbols and the first parity check symbol of the piggybacked stripes.
	\item Region C contains all the parity check symbols of the protected stripes.
	\item Region D contains the last $(r-1)$ parity check symbols of the piggybacked stripes.
\end{itemize}

Once a systematic node failure happens, the repair procedure is supposed to regenerate the $(s+p)$ missing symbols in Region A and B. Similar to RSR-II codes, the systematic symbols in Region B are self-sustaining: According to the MDS property, missing symbols in one row of Region B could be recovered by the surviving symbols in the other $k$ rows. As for the systematic symbols in Region A, piggybacking functions are constructed to protected them. These piggyback functions are supposed to be embedded in Region D. The size of Region D is $(r-1)p$, i.e., at most $(r-1)p$ piggyback functions can be designed. It is a noteworthy fact that the $s$ failed protected symbols in one row of Region A should be simultaneously recovered by solving a set of linear combinations. In order to guarantee that there are enough piggyback functions to simultaneously recover those $s$ missing symbols in Region A, the following inequality must be satisfied when we choose the parameters $s$ and $p$.
\begin{equation}
	(r-1)p\ge s.
\end{equation}
In the remainder this subsection, an method of the construction of $(r-1)p$ piggyback functions is illustrated as follows.

\begin{itemize}
	\item[1] Construct a $\lceil \frac{ks}{(r-1)p} \rceil\times (r-1)p$ empty piggybacking array.\\
	Each column of this piggybacking array corresponds to one piggyback function. 
	\item[2] Fill the protected symbols in Region A into the piggybacking array.\\
	The protected symbols in Region A form a $k\times s$ array as shown in Fig.\ref{fig.mul_stps}. Step 2 takes these symbols in rowwise from the $k\times s$ array and fills them into the piggybacking array. Obviously, if $ks$ is not divisible by $(r-1)p$, the last row of this piggyback array would not be full.
	\item[3] Obtain the $(r-1)p$ piggybacking functions, and add them in Region D.
	After all protected symbols are allocated into the piggyback array, sum the symbols in each column up. Thus, $(r-1)p$ piggybacking functions are obtained, and they can be added into Region D in an arbitrary order.
\end{itemize}
It is remarkable that the piggyback functions are only summations of some protected symbols. As a result, the recovery of missing protected symbols could be very simple. An example is presented to illustrate the partition method and the repair procedure.
\begin{example}
	Consider an $(8,4)$ systematic MDS code as the base code. Set $s=3$, and $p=2$. Denote $\ba,\bb,\bc,\bd,\be$ of length 4 as the 5 input message vectors. Thus, the original storage array is
	\begin{center}
		\small
		\begin{tabular}{|p{1.3em}p{1.3em}p{1.3em}|p{1.3em}p{1.3em}|}
			\hline
			$a_1$ & $b_1$ & $c_1$ & $d_1$ & $e_1$\\
			$a_2$ & $b_2$ & $c_2$ & $d_2$ & $e_2$\\
			$a_3$ & $b_3$ & $c_3$ & $d_3$ & $e_3$\\
			$a_4$ & $b_4$ & $c_4$ & $d_4$ & $e_4$\\
			\cline{1-3}
			$\bp_1^T\ba$&$\bp_1^T\bb$&$\bp_1^T\bc$&$\bp_1^T\bd$&$\bp_1^T\be$\\
			\cline{4-5}
			$\bp_2^T\ba$&$\bp_2^T\bb$&$\bp_2^T\bc$&$\bp_2^T\bd$&$\bp_2^T\be$\\
			$\bp_3^T\ba$&$\bp_3^T\bb$&$\bp_3^T\bc$&$\bp_3^T\bd$&$\bp_3^T\be$\\
			$\bp_4^T\ba$&$\bp_4^T\bb$&$\bp_4^T\bc$&$\bp_4^T\bd$&$\bp_4^T\be$\\
			\hline
		\end{tabular}			
	\end{center}
	The protected symbols in Region A are $\{a_1,a_2,a_3,a_4\}$, $\{b_1,b_2,b_3,b_4\}$, $\{c_1,c_2,c_3,c_4\}$ and $\{d_1,d_2,d_3,d_4\}$. Fill them into a $2\times6$ piggyback array. We have
	\begin{center}
		\small
		\begin{tabular}{cccccc}
			$a_1$ & $b_1$ & $c_1$ & $a_2$ & $b_2$ & $c_2$\\
			$a_3$ & $b_3$ & $c_3$ & $a_4$ & $b_4$ & $c_4$
		\end{tabular}			
	\end{center}
	Sum the symbols in each column up, and then we achieve the six piggyback functions $(a_1+a_3),(b_1+b_3),(c_1+c_3),(a_2+a+4),(b_2+b_4),(c_2+c_4)$. Finally, the generalized piggybacking code can be constructed as follows
	\begin{center}
		\small
		\begin{tabular}{|p{1.3em}p{1.3em}p{1.3em}|p{6.4em}p{6.3em}|}
			\hline
			$a_1$ & $b_1$ & $c_1$ & $d_1$ & $e_1$\\			
			$a_2$ & $b_2$ & $c_2$ & $d_2$ & $e_2$\\
			$a_3$ & $b_3$ & $c_3$ & $d_3$ & $e_3$\\
			$a_4$ & $b_4$ & $c_4$ & $d_4$ & $e_4$\\
			\cline{1-3}
			$\bp_1^T\ba$&$\bp_1^T\bb$&$\bp_1^T\bc$&$\bp_1^T\bd$&$\bp_1^T\be$\\
			\cline{4-5}
			$\bp_2^T\ba$&$\bp_2^T\bb$&$\bp_2^T\bc$&$\bp_2^T\bd+(a_1+a_3)$&$\bp_2^T\be+(b_1+b_3)$\\
			$\bp_3^T\ba$&$\bp_3^T\bb$&$\bp_3^T\bc$&$\bp_3^T\bd+(c_1+c_3)$&$\bp_3^T\be+(a_2+a_4)$\\
			$\bp_4^T\ba$&$\bp_4^T\bb$&$\bp_4^T\bc$&$\bp_4^T\bd+(b_2+b_4)$&$\bp_4^T\be+(c_2+c_4)$\\
			\hline
		\end{tabular}			
	\end{center}
\end{example}

\subsection{Analysis on repair bandwidth}\label{sec.analysis_PB}

Recall the construction of piggyback functions in Section.\ref{sec.desgn_generalized}. If $ks$ is not dividable by $(r-1)p$, the systematic symbols partitioned into the $(r-1)p$ piggyback functions are uneven. Here, we define the $(r-1)p$ sizes of these piggyback functions as the numbers of contained systematic symbols in Region A. Without loss of generality, assume the $(r-1)p$ sizes are not all the same, and denote them as $n_1,n_2,\cdots,n_{(r-1)p}$. Obviously, they satisfy that
\begin{equation}\label{equ.size_pb}
	\sum_{i=1}^{(r-1)p}n_i=ks.
\end{equation}

Suppose that the $l$-th systematic node fails, $l\in\{1,\cdots,k\}$. All remaining symbols stored in Region B except node $l$ are needed to reconstruct $\{a_{s+1,l},\cdots,a_{s+p,l}\}$ with the MDS property. The amount transmitted in this step is $kp$ symbols. In Region D, the $s$ parity check symbols containing the piggyback functions of $\{a_{1,l},\cdots,a_{s,l}\}$ are required to recover the $s$ missing protected symbols. Moreover, the components along $\{\ba_{s+1},\cdots,\ba_{s+p}\}$ should be subtracted out from the $s$ downloaded parity check symbols. However, the left piggybacking functions are still involved with some other protected symbols besides $\{a_{1,l},\cdots,a_{s,l}\}$. Hence, more symbols in Region A are needed. Assume the sizes of these $s$ piggybacking functions are $n_{i_1},n_{i_2},\cdots,n_{i_s}$. The download amount of systematic symbols from Region A in this step is $(n_{i_1}+n_{i_2}+\cdots+n_{i_s}-s)$.

Now we derive the total bandwidth of repairing all the $k$ systematic nodes. Symbols in Region B need to be downloaded $k^2p$ times. Consider a parity check symbol stored in Region D. Suppose the size of the piggybacking function embedded in this parity check symbol is $n_i$ $(i\in\{1,\cdots,(r-1)p\})$. During the repair procedures, the parity check symbol needs to be downloaded $n_i$ times. Meanwhile, each of the $n_i$ involved systematic symbols in Region A needs to be downloaded $(n_i-1)$ times. Therefore, the total repair bandwidth of all the $k$ systematic nodes is $k^2p+\sum_{i=1}^{(r-1)p}n_i^2$.

From the above, the average repair ratio $\gamma^{sys}_2$ is
\begin{equation}\label{equ.gamma2}
	\gamma^{sys}_2 = \frac{1}{k^2(s+p)}(k^2p+\sum\limits_{i=1}^{(r-1)p}n^2_i).
\end{equation}
Rewrite Equation (\ref{equ.gamma2}) as
\begin{eqnarray}
\gamma^{sys}_2&=&\frac{1}{k^2(s+p)}(k^2p+\sum\limits_{i=1}^{(r-1)p}n^2_i) \nonumber\\
&=&\frac{1}{k^2(s+p)}\Big[k^2p+\frac{(\sum\limits_{i=1}^{(r-1)p}n_i)^2+\sum\limits_{i\ne j}(n_i-n_j)^2}{(r-1)p}\Big] \nonumber\\
&=&\frac{1}{k^2(s+p)}\Big[k^2p+\frac{k^2s^2+\sum\limits_{i\ne j}(n_i-n_j)^2}{(r-1)p}\Big].
\end{eqnarray}
Without loss of generality, assume $ks$ is not dividable by $(r-1)p$, and
\begin{equation}
	t'_l=\Bigl\lfloor \frac{ks}{(r-1)p}\Bigr\rfloor,\ t'_h=\Bigl\lceil \frac{ks}{(r-1)p}\Bigr\rceil,\ t'=ks-t'_l(r-1)p.
\end{equation}
Thus, $t'$ out of $(r-1)p$ piggyback functions have the size of $t'_h$, and the rest $(r-1)p-t'$ ones have the size of $t'_l$. Then, $\gamma^{sys}_2$ goes to
\begin{equation}
	\gamma^{sys}_2=\frac{1}{k^2(s+p)}\Big[k^2p+\frac{k^2s^2}{(r-1)p}+\frac{t'((r-1)p-t')}{(r-1)p}\Big].
\end{equation}

In a DSS, the parameters of base code $(k,r)$ are given. Thus, $\gamma^{sys}_2$ is varied with different values of $(s,p)$. In order to explore the relationship between $\gamma^{sys}_2$ and the proportion of protected instances $p_p=\frac{s}{s+p}$, the lower and upper bounds of $\gamma^{sys}_2$ are derived as follows,
\begin{eqnarray}
	\gamma^{sys}_2 &\ge& \frac{1}{k^2(s+p)}(k^2p+\frac{k^2s^2}{(r-1)p})\nonumber\\
	&=& \frac{p}{s+p}+\frac{s}{s+p}\cdot\frac{s}{(r-1)p}\\
	\gamma^{sys}_2 &\le& \frac{1}{k^2(s+p)}\Big[k^2p+\frac{k^2s^2}{(r-1)p}+\frac{(r-1)p}{4}\Big]\nonumber\\
	&=& \frac{p}{s+p}\Big(1+\frac{r-1}{4k^2}\Big)+\frac{s}{s+p}\cdot\frac{s}{(r-1)p}
\end{eqnarray}

Rewrite the lower and upper bounds as functions $\Gamma_{low}(p_p)$ and $\Gamma_{up}(p_p)$ of $p_p$. Then,
\begin{eqnarray}
	\Gamma_{low}(p_p)&=&(1-p_p)+\frac{{p_p}^2}{1-p_p}\cdot\frac{1}{r-1}\\
	\Gamma_{up}(p_p)&=&(1-p_p)\Big(1+\frac{r-1}{4k^2}\Big)+\frac{{p_p}^2}{1-p_p}\cdot\frac{1}{r-1}
\end{eqnarray} 

\begin{example}
	Assume the code rate of the base code is $0.5$, i.e., $k=r$. For various $r$'s, Figure \ref{fig.gamma_with_pp} shows the curves of $\Gamma_{low}(p_p)$ and $\Gamma_{up}(p_p)$ with $p_p$.
	\begin{figure}
		\centering
		\includegraphics[width=7.5cm]{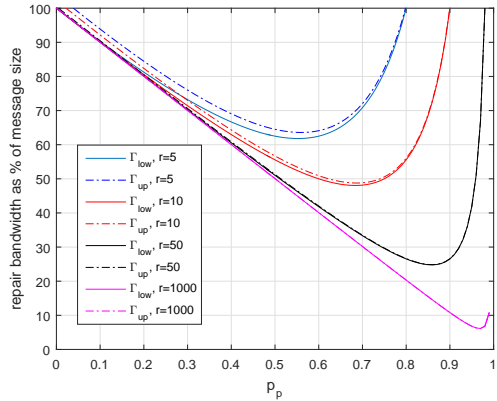}
		\caption{The lower and upper bounds with various $p_p$.}\label{fig.gamma_with_pp}
	\end{figure}
	
	It illustrates that the lower bound $\Gamma_{low}(p_p)$ and upper bound $\Gamma_{up}(p_p)$ are close to each other. Moreover, both of them can reach their extreme points by optimizing $p_p$ which implies that the generalized piggybacking code can obtain optimum $\gamma^{sys}_2$ with appropriate parameters $(s,p)$. 
\end{example}

\begin{figure}
	\centering
	\includegraphics[width=7.5cm]{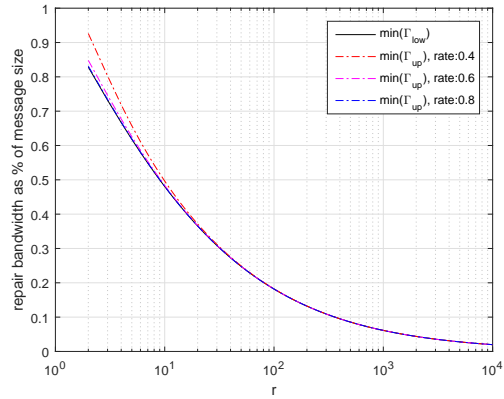}
	\caption{Minimum values of $\gamma^{sys}_1$ and $\gamma^{sys}_2$.}\label{fig.gamma_curve}
\end{figure}

Further analyze the optimum condition for $\gamma^{sys}_2$ with the derivatives of $\Gamma_{low}(p_p)$ and $\Gamma_{up}(p_p),$ which are with respect to $p_p$ and listed as follows
\begin{eqnarray}
	\frac{\partial\Gamma_{low}(p_p)}{\partial p_p}&=&\frac{-rp_p^2+2rp_p-(r-1)}{(r-1)(1-p_p)^2}\\
	\frac{\partial\Gamma_{up}(p_p)}{\partial p_p}&=&\frac{-rp_p^2+2rp_p-(r-1)}{(r-1)(1-p_p)^2}-\frac{r-1}{4k^2}.
\end{eqnarray}
Let $\frac{\partial\Gamma_{low}(p_p)}{\partial p_p}$ and $\frac{\partial\Gamma_{up}(p_p)}{\partial p_p}$ equal to zero. Then, we work out the minimum values of $\Gamma_{low}(p_p)$ and $\Gamma_{up}(p_p)$ as follows,
\begin{enumerate}
	\item $\min(\Gamma_{low}(p_p))=\frac{2}{\sqrt{r}+1}$, when $p_p=1-\frac{1}{\sqrt{r}}$;
	\item $\min(\Gamma_{up}(p_p))=\frac{-2+2\sqrt{r+\frac{(r-1)^2}{4k^2}}}{r-1}$, when $p_p=1-\frac{1}{\sqrt{r+\frac{(r-1)^2}{4k^2}}}$.
\end{enumerate}
The results indicate that 
\begin{enumerate}
	\item $\min(\Gamma_{low}(p_p))$ is only determined by the number of parity check nodes $r$;
	\item $\min(\Gamma_{up}(p_p))$ is determined by both $k$ and $r$. However, for high code rate, $\min(\Gamma_{up}(p_p))$ is dominantly determined by $r$;
	\item $\min(\Gamma_{up}(p_p))$ corresponds closely to $\min(\Gamma_{low}(p_p))$. In other words, there exists a generalized piggybacking code whose repair ratio is very close to the lower bound.
\end{enumerate}
Figure \ref{fig.gamma_curve} shows the curves of $\min(\Gamma_{low}(p_p))$ and $\min(\Gamma_{up}(p_p))$ with $r$.
It implies that
\begin{equation}
	\min(\gamma^{sys}_2)\approx\min(\Gamma_{low}(p_p))=\frac{2}{\sqrt{r}+1}.
\end{equation}

\begin{figure}
	\centering
	\includegraphics[width=7.5cm]{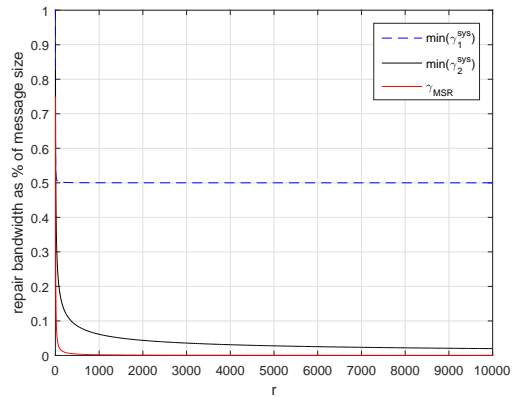}
	\caption{Lower bounds on the average repair bandwidths.}\label{fig.comparison}
\end{figure}

At the end of this subsection, we perform asymptotic analyses of $\min{(\gamma^{sys}_1)}$ and $\min{(\gamma^{sys}_2)}$, and compare them with the repair ratio of \emph{minimum storage regenerating (MSR)} codes $\gamma_{MSR}$. The limits of $\min{(\gamma^{sys}_1)}$ and $\min{(\gamma^{sys}_2)}$ as $r$ approaches infinity are
\begin{eqnarray}
\lim\limits_{r\to+\infty}{\min{(\gamma^{sys}_1)}} &=& \lim\limits_{r\to+\infty}{\frac{r-1}{2r-3}}=0.5\\
\lim\limits_{r\to+\infty}{\min{(\gamma^{sys}_2)}} &=& \lim\limits_{r\to+\infty}{\frac{2}{\sqrt{r}+1}}=0.
\end{eqnarray}
As described in \cite{regeneratingcode2010,dimakis2011survey,yang2015pbmsr}, MSR codes which correspond to the best storage efficiency are one of two most important classes of regenerating codes. The repair bandwidth for one failure node is
\begin{equation}
	B_{MSR}=\frac{{\cal M}d}{k(d-k+1)},
\end{equation}
where ${\cal M}$ represents the size of original messages, $d$ denotes the number of accessed surviving nodes, and $k$ is the dimension of the MSR code. For the sake of simple comparison, we set the code rate to $0.5$, and $d=n-1$ such that the MSR code provides the highest repair efficiency. Thus,
\begin{equation}
	\gamma_{MSR}=\frac{B_{MSR}}{{\cal M}}=\frac{2}{r}-\frac{1}{r^2}.
\end{equation}
The curves of $\min{(\gamma^{sys}_1)}$, $\min{(\gamma^{sys}_2)}$ and $\gamma_{MSR}$ are shown in Fig.\ref{fig.comparison}.
It shows that $\min{(\gamma^{sys}_2)}$ approaches zero instead of $50\%$ as the number of parity check nodes tends to infinity. As a result, compared with RSR-II codes,  generalized piggybacking codes are able to provide more efficient node repair with less bandwidth. Moreover, $\min{(\gamma^{sys}_2)}$ is closer to $\gamma_{MSR}$ - the theoretical lower bound of repair ratio.

Table \ref{tab.efficy_comp} compares the repair efficiency of RSR-II codes and generalized piggybacking codes with various code parameters $n$ and $k$. It is illustrated that with the increasing of the number of parity check nodes, generalized piggybacking codes can reach smaller repair bandwidth.
\begin{table}
	\caption{Efficiency comparison for different explicit codes}\label{tab.efficy_comp}
	\centering
	\begin{tabular}{|c|c|c|c|c|c|}
		\hline
		\multirow{2}{*}{$n,k$} & \multicolumn{2}{|c|}{RSR-II codes} & \multicolumn{3}{|c|}{generalized piggybacking codes} \\
		\cline{2-6}
		& stripes & $\gamma^{sys}_1$ & $s,p$ & stripes & $\gamma^{sys}_2$ \\
		\hline
		$10,5$ &  7 & 0.5886  & $1,1$ & 2 & 0.6400\\
		\hline		
		$20,10$ & 17 & 0.5341 & $2,1$ & 3 & 0.4867\\
		\hline
		$30,15$ & 27 & 0.5207 & $3,1$ & 4 & 0.4133\\
		\hline
		$40,20$ & 37 & 0.5147 & $4,1$ & 5 & 0.3700\\
		\hline
		$50,25$ & 47 & 0.5114 & $4,1$ & 5 & 0.3344\\
		\hline
		$80,40$ & 77 & 0.5068 & $5,1$ & 6 & 0.2740\\
		\hline
		$200,100$ & 197 & 0.5026 & $9,1$ & 10 & 0.1819\\
		\hline
	\end{tabular}
\end{table}

\subsection{Analysis on decoding complexity}

In this subsection, the complexity of node repair procedure of generalized piggybacking codes is analyzed first. Then the comparison with RSR-II codes is performed. 
It is shown that the computational complexity for repairing a single systematic node cost by generalized piggybacking codes is much less than that of RSR-II codes.
 
As the statement in Section.\ref{sec.analysis_rsrii} and \ref{sec.analysis_PB}, piggybacking codes adopt two kinds of calculations to repair a failed node. MDS decoding is used for the recovery of the missing symbols in non-protected or piggybacked stripes, while solving linear combinations is employed to reconstruct the missing symbols in protected stripes. Recall the generalized piggybacking code, in Section.\ref{sec.desgn_generalized}, which has $s$ protected stripes and $p$ piggybacked stripes. The repair procedure of the $l$-th systematic node is described in Section.\ref{sec.analysis_PB}.

In order to recover $a_{s+i,l}$ - the missing symbol of the $i$-th piggybacked stripe, the symbols $\{a_{s+i,1},\cdots,a_{s+i,l-1},a_{s+i,l+1},\cdots,a_{s+i,k},\bp^T_1\ba_{s+i}\}$ are required. Denote the vector representation of $\bp_j$ ($j\in\{1,\cdots,r\}$) as $[p_{j,1},\cdots,p_{j,k}]$. Then, $a_{s+i,l}$ can be worked out by the below equation.
\begin{eqnarray}
	a_{s+i,l}&=&p_{j,l}^{-1}[\bp^T_1\ba_{s+i}-(a_{s+i,1}p_{j,1}+\cdots+\nonumber\\
	&&a_{s+i,l-1}p_{j,l-1}+a_{s+i,l+1}p_{j,l+1}+\cdots+\nonumber\\
	&&a_{s+i,k}p_{j,k})].
\end{eqnarray}
Hence, the MDS decoding for the recovery of one missing symbol in a piggybacked stripe costs $k$ multiplications and $(k-1)$ additions.

Consider the recovery of $a_{i,l}$ - the missing symbol in the $i$-th protected stripe. According to the description of Section.\ref{sec.desgn_generalized}, we denote the piggyback function which involves $a_{i,l}$ together with other $(n_x-1)$ protected symbols as $F_x$. In order to reconstruct $a_{i,l}$, from Region D, the stored symbol $\xi$ containing $F_x$ is needed, and the $(n_x-1)$ surviving protected symbols are also required. Hence, $a_{i,l}$ can be figured out as follows.
\begin{itemize}
	\item Compute the parity check symbol in $\xi$. This step costs $k$ multiplications and $(k-1)$ additions.
	\item Subtract the parity check symbol from $\xi$. Thus, 1 addition is needed.
	\item Subtract the $(n_x-1)$ surviving protected symbols form the left $F_x$. Thus, $(n_x-1)$ additions are required.	
\end{itemize}
Actually, $n_x$ represents the size of the piggyback function $F_x$, i.e., $n_x$ equals to $t'_l$ or $t'_h$. Therefore, solving linear combinations for one missing protected symbol costs $k$ multiplications and $\frac{ks}{(r-1)p}+k-1$ additions, on average.

The computational complexity of MDS decoding and solving linear combinations is listed in Table \ref{tab.pb_cplx}.
\begin{table}
	\caption{Computational complexity for node repair}\label{tab.pb_cplx}
	\centering
	\begin{tabular}{|c|c|c|}
		\hline
		& Multiplications & Additions\\
		\hline
		MDS decoding & $k$ & $(k-1)$ \\
		\hline
		Solving linear & \multirow{2}{*}{$k$} & \multirow{2}{*}{$\frac{ks}{(r-1)p}+k-1$}\\
		combinations & &\\
		\hline
	\end{tabular}
\end{table}

According the analysis in Section.\ref{sec.analysis_rsrii}, solving linear combinations is introduced by piggybacking codes to reduce the repair bandwidth of partial missing symbols. For RSR-II codes, $(r-1)$ missing protected symbols need to be simultaneously recovered by solving a group of $(r-1)$ linear functions. As a result, we have to perform Gaussian elimination. However, for generalized piggybacking codes, piggyback functions are simple summations of some protected symbols. Compared with the calculations for MDS decoding, those for solving linear combinations cost only $\frac{ks}{(r-1)p}$ more additions. Thus, the generalized piggybacking framework is able to provide high repair efficiency because it can significantly reduce the repair bandwidth for a single failed systematic node with low computational complexity.

\section{Conclusion and Discussion}
This paper presents a generalized piggybacking construction with various protected instances and piggybacked instances. Compared with the previous design, our proposed generalized piggybacking codes can save more repair bandwidth by optimizing the proportion of protected instances. When the number of parity check nodes tends to infinity, the average repair bandwidth as a fraction of total messages approaches zero. Moreover, complexity analysis demonstrates that generalized piggybacking codes are able to efficiently repair the failed node with reasonable complexity overhead.

In fact, if we look at piggybacking functions from the view of error-correction codes, piggybacking codes are perfect encounter between codes with small minimum Hamming distance and codes with large minimum Hamming distance. The repair of systematic symbols in piggybacked stripes is relied on the base codes of these stripes. These base codes have strong erasure-correction capability due to their large minimum distance. However, it results in strong correlation among all the symbols. Thus, decoding of these ‘good codes’ requests large amount of data access. For the repair of protected stripes, piggybacking functions are linear combinations of the
protected systematic symbols. In other words, these symbols together with piggyback functions can be considered as linear codes with small minimum distance.
Since these ‘bad codes’ have weak correlation among symbols, their decoding requests small amount of data access.

\section{Acknowledgment}
We sincerely thank Prof. Shu Lin and Dr. Zhiying Wang for their constructive suggests. This paper received funding from NSAF under Grant U1530117 and National Natural Science Foundation of China under Grant 61471022, and also sponsored by Laboratory Independent Innovation project of Qian Xuesen Laboratory of Space Technology.


\end{document}